\documentclass[aps,pra]{revtex4}%,preprint
\begin{document}

\title{Comment on quant-ph/0506105: The modified Grover algorithm cannot improve the Grover algorithm}
\author{Gui Lu Long$^{1,2}$ }
\affiliation{ $^1$ Key Laboratory For Quantum Information and
Measurements and Department of Physics, Tsinghua University,
Beijing 100084, China\\
$^2$ Key Laboratory for Atomic and Molecular NanoSciences,
Tsinghua University, Beijing 100084, China}
\date{\today }
\maketitle

In a recent paper in this eprint server \cite{r1}, A S Gupta, M.
Gupta and A. Pathak proposed a modified Grover algorithm that
would exponentially accelerate the unsorted database search
problem if the number of marked items is known. If this were true,
it would represent a major fundamental breakthrough in computer
science, mathematics, quantum information and other related
branches of sciences.

However the algorithm is not valid. We will explain it in this
brief comment.

The very reason that their modified algorithm can not be true is
the realization of  Eq. (6) in their paper. This unitary operation
actually implement a $k\theta$ rotation in a 2-dimensional space
which will require $k/2\simeq O(\sqrt{N})$ iterations in the
Grover algorithm. This operation is central to the exponential
speedup. However, one cannot build this operation without the use
of the query. When one uses the query, it would require
$O(\sqrt{N})$ number of queries to implement this operation. Hence
the exponential speedup is only nominal.

It has been shown that Grover's algorithm is optimal\cite{z}. This
optimality theorem has been proven valid sofar. Moreover, it seems
that classical NP-Complete problems such as the SAT problem is
unlikely to find an exponential speedup, even with the use of a
quantum computer.

\end{document}